\documentclass[12pt,preprint]{aastex}
\usepackage{graphicx}

\slugcomment{Submitted to the Astrophysical Journal Letters}
\shorttitle{Grain-Depleted Accretion on T~Tauri Stars}
\shortauthors{J.J.~Drake et al.}
\begin{document}

\title{X-ray Diagnostics of Grain Depletion in Matter Accreting onto
T~Tauri Stars}

\author{Jeremy J.~Drake\altaffilmark{1},
Paola~Testa\altaffilmark{2},
Lee Hartmann\altaffilmark{1}}
\affil{$^1$Smithsonian Astrophysical Observatory,
MS-3, \\ 60 Garden Street, \\ Cambridge, MA 02138}
\email{jdrake@cfa.harvard.edu}
\affil{$^2$Kavli Institute for Astrophysics and Space Research, 
Massachusetts Institute for Technology,\\
70 Vassar Street,Cambridge, MA 02139}

\begin{abstract}
Recent analysis of high resolution {\it Chandra} X-ray spectra has
shown that the Ne/O abundance ratio is remarkably constant in stellar
coronae.  Based on this result, we point out the utility of the Ne/O
ratio as a discriminant for accretion-related X-rays from T~Tauri
stars, and for probing the measure of grain-depletion of the accreting
material in the inner disk.  We apply the Ne/O diagnostic to the
classical T~Tauri stars BP~Tau and TW~Hya---the two stars found to
date whose X-ray emission appears to originate, at least in part, from
accretion activity.  We show that TW~Hya appears to be accreting
material which is significantly depleted in O relative to Ne.  In
constrast, BP~Tau has an Ne/O abundance ratio consistent with that
observed for post-T~Tauri stars.  We interpret this result in terms of
the different ages and evolutionary states of the circumstellar disks
of these stars.  In the young BP~Tau disk (age $\sim 0.6$~Myr) dust is
still present near the disk corotation radius and can be ionized and
accreted, re-releasing elements depleted onto grains.  In the more
evolved TW~Hya disk (age $\sim 10$~Myr), evidence points to ongoing
coagulation of grains into much larger bodies, and possibly planets, 
that can resist the drag
of inward-migrating gas, and accreting gas is consequently depleted of
grain-forming elements.
\end{abstract}


\section{Introduction}
\label{s:intro}

X-rays, at least in principle, present a powerful method of studying
directly the accretion processes of star formation.  In the case of
classical T~Tauri stars (CTTS), the magnetospheric accretion paradigm
posits that material is funnelled by magnetic fields onto the star
from a truncated disk \citep[e.g.\
][]{Uchida.Shibata:84,Bertout.etal:88}, rather than through direct
surface interaction with the disk itself.  Matter impacts the stellar
surface at free-fall velocities of up to a few hundred km~s$^{-1}$.
The accreting material is then expected to form a shock at the stellar
surface.  The resulting shocked plasma temperature is of order a few
million K, and it will therefore radiate predominantly in X-rays.  For
the ranges of mass accretion rates inferred for CTTS of
$10^{-6}$-$10^{-10}M_\odot$~yr$^{-1}$ \citep[e.g.\
][]{Hartigan.etal:95,Gullbring.etal:98,Johns-Krull.etal:00}, the X-ray
luminosities resulting from accretion should be $\ga
10^{31}$~erg~cm$^{-2}$~s$^{-1}$---easily sufficient to be observed in
nearby associations of T~Tauri stars and regions of star formation. 

The above scenario of copious accretion-driven X-rays from CTTS
contrasts sharply with what is observed.  While strong X-ray emission
appears to be a ubiquitous characteristic of CTTS \citep[e.g.\
][]{Feigelson.Montmerle:99}, the source plasma temperatures are an
order of magnitude higher than can be produced in accretion shocks and
heating can instead be attributed to magnetic processes analogous to
coronal activity in late-type main sequence and more evolved stars.
Accretion activity is largely revealed by strong UV-optical continuum
emission, suggesting that accretion shocks are either formed too deep
in the stellar atmosphere to be observed \citep{Drake:05}, or that
infall velocities are insufficient to attain X-ray temperatures in the
shock.  Nevertheless, two stars now stand
out as probable exceptions and examples of objects whose high
resolution X-ray spectra appear to be produced, at least in part, by
accretion: TW~Hya \citep{Kastner.etal:02,Stelzer.Schmitt:04} and
BP~Tau \citep{Schmitt.etal:05}.

TW~Hya has a dominant plasma temperature of $3\times 10^6$~K, as
expected from an accretion shock resulting from free-fall of gas from
a truncated disk.  Based on the density-sensitive
He-like Ne and O lines, both TW~Hya and BP~Tau seem to be
characterised by plasma with high electron densities $n_e\sim
10^{11}$-$10^{13}$~cm$^{-3}$ at temperatures of $\sim 3\times 10^6$~K
\citep{Kastner.etal:02,Stelzer.Schmitt:04}, 
in contrast to all other single and active binary stars studied in
the surveys of \citet{Testa.etal:04} and \citet{Ness.etal:04}, where
He-like O lines indicate $n_e\sim 10^{10}$~cm$^{-3}$.  The X-ray
spectra of both stars exhibit extremely weak lines of Mg, Si and Fe,
and instead are dominated by O and Ne.  While this pattern is
reminiscent of that seen in very active RS~CVn-type binaries
\citep[e.g.\ ][]{Huenemoerder.etal:01}, \cite{Stelzer.Schmitt:04}
echoed the earlier suggestion of \citet{Herczeg.etal:02} that the
metal depletion is instead a signature of the accretion of
grain-depleted gas.  If the latter is indeed the case, X-rays then
provide a unique means of investigating the matter in the inner
accretion disk that is in the process of accreting.

In this {\it Letter}, we draw on the recent findings of
\citet{Drake.Testa:05}, based on {\it Chandra} High Energy
Transmission Grating (HETG) spectra of nearby stars, that the Ne/O
abundance ratio is remarkably constant in stellar coronae.  We use
this result to investigate Ne/O in TW~Hya and BP~Tau, and show that
this ratio provides a diagnostic of the depletion of O in grains in
the circumstellar disk, and in particular in the very inner disk from
which accreting material derives.


\section{Observations and Analysis}
\label{s:obs}

We use spectral line fluxes for the H-like and He-like resonance lines
of Ne and O obtained from {\it XMM-Newton} observations of TW~Hya
(acquired 2001 July) and BP~Tau (acquired 2004 August) by
\cite{Stelzer.Schmitt:04} and \cite{Schmitt.etal:05}, respectively;
the reader is referred to these works for further details regarding
the observations.  To provide an additional TW~Hya association
comparison for TW~Hya itself, we also analysed the same lines for the
CTTS TWA~5 studied recently by \citet{Argiroffi.etal:05} based on {\it
XMM-Newton} spectra.  

TW~Hya was observed earlier by the {\it Chandra} HETG and ACIS-S, on
2000 July 18 at UT10:19 \citep[see][for further
details]{Kastner.etal:02}.  Here, we have analysed the products of
standard CIAO 3.0 processing, and in particular the Medium Energy
Grating spectrum.  The measurement of Ne and O line spectral line
fluxes and all calculations were performed using the IDL Package for
INTeractive Analysis of Line Emission (PINTofALE)\footnote{PINTofALE
is freely available from http://hea-www.harvard.edu/PINTofALE/ }
following the methods described by \citet{Testa.etal:04}.  The line
fluxes of interest for this study are listed in Table~\ref{t:flx}.

The conversion of Ne and O fluxes into the Ne/O abundance ratio by
number, $A_{Ne}/A_{O}$, is described by
\citet{Drake.Testa:05}.  Briefly, an abundance ratio in an 
optically-thin, collision-dominated plasma can be
derived by using lines whose contribution functions $G_{ji}(T)$---the
product of the parent ion population and the line emissivity---have
very similar temperature dependence, simply from the ratio of observed
line fluxes (corrected for attenuation by interstellar extinction), 
$F_O$ and $F_{Ne}$:
\begin{equation}
\frac{A_{Ne}}{A_{O}}=
\overline{\left(\frac{G_O}{G_{Ne}}\right)}\frac{F_{Ne}}{F_{O}}
\end{equation}
For Ne and O, such a ratio of $G_{ji}(T)$ functions can be constructed
from the O~VIII, Ne~IX and Ne~X resonance lines $1s2p\, ^1P_1
\rightarrow 1s^2\, ^1S_0$ and $2p\, ^2P_{3/2,1/2} \rightarrow 1s\,
^2S_{1/2}$, combined as follows
\begin{equation}
\overline{\left(\frac{G_O}{G_{Ne}}\right)}=
\frac{G_{OVIII}}{G_{NeIX}+0.15G_{NeX}}.
\end{equation}  
\citet{Drake.Testa:05} found $\overline{G_{O}/G_{Ne}}=1.2\pm 0.1$ for
active stars. 
Ne/O abundance ratios derived for BP~Tau, TW~Hya and TWA~5
using this method are also listed in
Table~\ref{t:flx}, together with the intervening absorbing columns
used to correct the observed line fluxes for attenuation. 
These Ne/O ratios are compared with those
presented by \citet{Drake.Testa:05} for a sample of 21 post-T~Tauri
stars, including single main-sequence stars, giants, and
tidally-interacting binaries in Figure~\ref{f:ne_o}.

\section{Discussion}
\label{s:discuss}

\begin{figure}
\epsscale{1.0}
\plotone{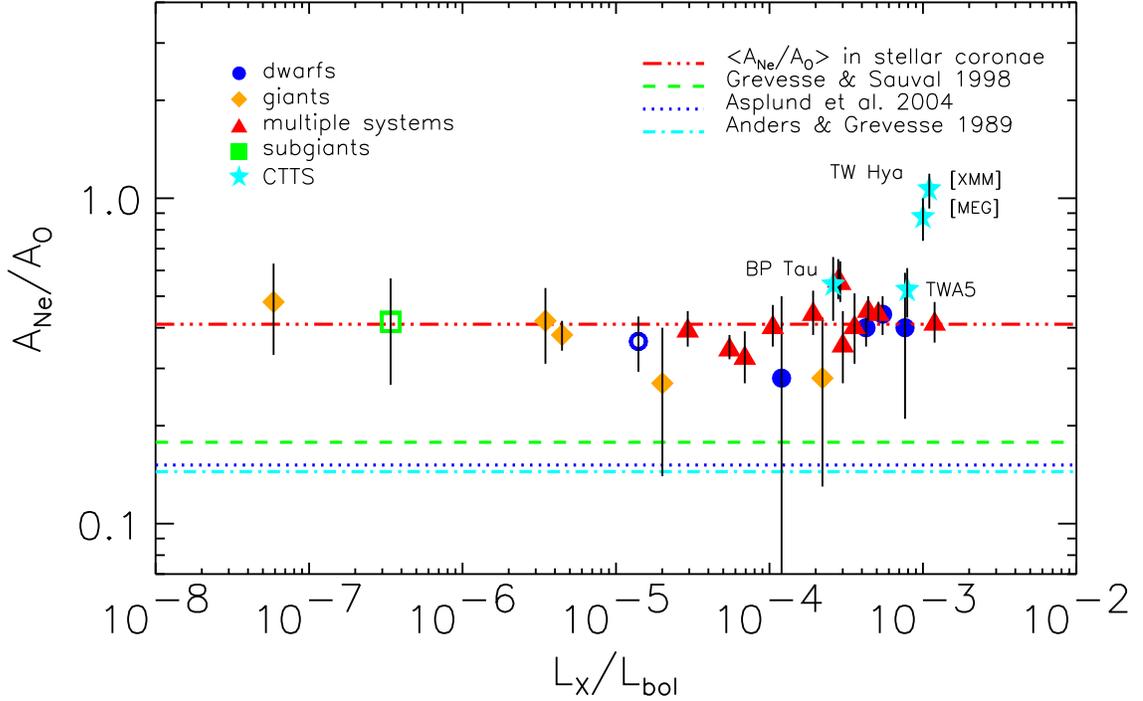}
\caption{
Derived Ne/O abundance ratios by number as a function of the
coronal activity index $L_X/L_{bol}$.  Data points represent the
$A_{Ne}/A_{O}$ for the objects in the \citet{Drake.Testa:05} sample,
together with TW~Hya, BP~Tau and TWA~5 analysed here.  Error bars are based
on the quadrature addition of the uncertainties of line flux
measurement.  The other hollow symbols correspond to the stars Procyon
(F5~IV) and $\epsilon$~Eri (K2~V) for which the abundances were taken
from the literature \citep{Sanz-Forcada.etal:04} to better represent the
lower ranges of coronal activity.  The error-weighted mean Ne/O
abundance ratio obtained is $A_{Ne}/A_{O}=0.41$, or 2.7 times the
currently assessed value \citep{Asplund.etal:04} which is illustrated by
the dotted horizontal line.  The recommended values from earlier
studies in common usage
\citep{Grevesse.Sauval:98,Anders.Grevesse:89} are also illustrated.  }
\label{f:ne_o}
\end{figure}

Based on the Ne/O ratios found for post-T~Tauri stars,
\citet{Drake.Testa:05} drew two conclusions: (i) Ne/O is essentially
constant in stellar coronae, and in full-disk observations is not
susceptible to fractionation effects that often appear to characterise other
elements with lower first ionisation potentials (FIPs) 
\citep[e.g.\ ][]{Drake:03b}; (ii) the Ne/O abundance ratio in stars is
significantly larger than current assessments of the solar ratio (also
illustrated in Figure~\ref{f:ne_o}), but is in-line with inference
from solar oscillations \citep{Antia.Basu:05,Bahcall.etal:05}. 
The constancy of the coronal Ne/O ratio then allows us to use these
elements for diagnostics purposes.  Firstly, we can determine
whether or not the suspected accretion shock has an Ne/O ratio
consistent with that of the underlying star, as represented by the
constant coronal Ne/O ratio illustrated in Figure~\ref{f:ne_o}.  A
ratio significantly different to the coronal value would provide
further evidence that these lines are not formed in a ``normal''
coronal plasma.  Secondly, departures from the coronal Ne/O ratio can then
be used to infer compositional peculiarities in the accreting gas.

We conclude from Figure~\ref{f:ne_o} that the Ne/O ratio of
$A_{Ne}/A_{O}\sim 1.0$ we find in TW~Hya---in good agreement with the
value obtained earlier by \citet{Kastner.etal:02} based on a
differential emission measure analysis---is significantly higher than
that of more evolved stars (Figure~\ref{f:ne_o}).  In contrast, Ne/O
in BP~Tau is perfectly consistent with that found for the rest of the
sample of \citet{Drake.Testa:05}.  This latter result also supports
the interpretation of the Ne/O abundance in stellar coronae as
representative of the ambient local cosmos: under the assumption that
the Ne and O lines in BP~Tau are formed from accreting material, as
the anomalous O~VII line ratios analysed by \citet{Schmitt.etal:05}
suggest, the ratio seen in BP~Tau indicates that its circumstellar
material shares the same Ne/O ratio.

We can confirm that the Ne/O ratio found for TW~Hya does not simply a
reflect an anomalous local environmental composition by comparison
with the ratio seen in TWA~5, a CTTS from the same association as
TW~Hya.  The X-ray spectrum of TWA~5 has been analysed recently by
\citet{Argiroffi.etal:05}: unlike TW~Hya and BP~Tau, it does not show
the high plasma density signature of an accretion shock, and its Ne
and O lines can be interpreted as being formed in magnetically-heated
coronal plasma.  Using the \citet{Argiroffi.etal:05} O and Ne line
fluxes (Table~\ref{t:flx}) we obtain $A_{Ne}/A_{O}=0.52 \pm 0.09$---in
good agreement with their result based on a differential emission
measure analysis and in excellent agreement with that of the rest of
our star sample, except for TW~Hya.  Similarly,
\citet{Kastner.etal:04} find $A_{Ne}/A_{O}\sim 0.5$ for the TW Hya
association multiple, non-accreting, weak-lined T Tauri star system
HD~98800 based on {\it Chandra} HETG spectra. 

\citet{Herczeg.etal:02} suggested that the lack of Si in the UV
spectrum of the accretion shock of TW~Hya was a signature of the
accretion of grain-depleted material.  \citet{Stelzer.Schmitt:04}
invoked their suggestion to explain the metal-poor X-ray spectrum.
However, since tidally-interacting binaries have also been shown to
exhibit similar metal paucity \citep[see, e.g.\ reviews
by][]{Drake:03b,Audard:05}, it is not immediately clear that the
composition of the hot plasma in TW~Hya is significantly different.
The Ne/O ratio provides this evidence, and supports the conjecture
that TW~Hya is accreting grain-depleted gas.  Unlike comparisons of Ne
with metals such as Mg, Si and Fe, the Ne/O diagnostic appears to be
robust to the effects of compositional fractionation seen in coronal
plasma.  Moreover, the Ne and O lines are formed at the temperature of
$\sim 3\times 10^6$~K expected of the accretion shocks on T~Tauri
stars, whereas dominant lines of Fe XVII and higher charge states, Mg
and Si are all formed at higher temperatures.

In the context of shock-produced versus coronal X-rays, it is worth
noting that the observed departure from a ``normal'' chemical
composition cannot be mimiced by a corona whose plasma is fed by
accretion.  The convective turnover time for CTTS is of order a few
hundred days \citep{Gilliland:86,Kim.Demarque:96} and so accreting
plasma, and any abundance peculiarity in these last accumulating
fractions, is rapidly subsumed by the underlying star that would feed
the corona.

We interpret the different Ne/O ratios in BP~Tau and TW~Hya as arising
from different depletions of O in the accreting gas.  The comparison
of BP~Tau with post-T~Tauri stars indicates that it is accreting
material which is essentially undepleted.  In the case of TW~Hya, the
O depletion amounts to a factor of $\sim 2$ or more.  Why should
BP~Tau and TW~Hya exhibit distinctly different O depletion in their
accreting plasma?  We propose that this reflects the different ages
and evolutionary states of these stars.

There is some controversy concerning the age of BP~Tau, though this is
entirely the result of its uncertain Hipparcos parallax.  The
Hipparcos data indicate a distance of only 42-70~pc ($\pm
1\sigma$~range)---well in front of the Taurus cloud at 140~pc in which
it is generally thought to reside
\citep{Wichmann.etal:98,Favata.etal:98,Bertout.etal:99}.  Based on
this distance, \citet{Favata.etal:98} estimated BP~Tau to be as old as
35~Myr through comparison with evolutionary tracks.  However, both
\citet{Wichmann.etal:98} and \citet{Bertout.etal:99} refute the
Hipparcos distance based on its low statistical significance.
\citet{Bertout.etal:99} find that the astrometric fit to the Hipparcos
data for the distance of the Taurus cloud of 140~pc, instead of
42-70~pc, represents only a $2\sigma$ deviation, and point out that
BP~Tau is rather faint for the nominal distance to be reliable.  They
also show the Hipparcos distance of 32-52~pc for for DF Tau---the
other Taurus star found by \citet{Favata.etal:98} to be of anomalous
proximity---to be highly uncertain owing to binarity and photometric
variability.  Allowing for a somewhat smaller distance than 140~pc, but
not as extreme as the Hipparcos value, \citet{Simon.etal:00} estimate
a range of 2-10~Myr.  However, all other observational evidence points
to BP~Tau lying in the Taurus cloud, including radial velocity and
proper motion \citep{Hartmann.Stauffer:89,Bertout.etal:99}, extinction
of $A_V \sim 0.5$ \citep[e.g.\ ][and references
therein]{Gullbring.etal:96b,Muzerolle.etal:03} and strong interstellar
Na I absorption \citep{Gullbring.etal:96}.  \citet{Gullbring.etal:98}
estimate the age to be 0.6~Myr.

In contrast, the distance and age of TW~Hya are much more secure: with
an age of $\sim 10$~Myr \citep[e.g.\ ][]{Zuckerman.Song:04}, TW~Hya appears
significantly older than BP~Tau.  TW~Hya is in fact one of the oldest
CTTS known to be still accreting.  More importantly, near infra-red
measurements also show the inner disks of the two stars to be quite
different.  Veiling measurements for BP~Tau show a very large excess
of 0.6 at $2.2\mu$m \citep{Muzerolle.etal:03}, and $K-L=0.57\pm 0.23$
\citep{Kenyon.Hartmann:95}---typical of the accreting stars in Taurus,
95\%\ of which have $K-L$ colours in the range 0.35-1.2
\citep{Meyer.etal:97}.  \citet{Muzerolle.etal:03} find that BP~Tau has
an optically thick disk whose dust truncation radius is essentially at
the corotation radius from which magnetospheric accretion is thought
to take place---material from both gas and dust will therefore accrete
and this material will not be depleted.


Instead, TW~Hya has very little near-infrared excess shortward of
10$\mu$m, with $K-L=0.25$
\citep{Sitko.etal:00,Calvet.etal:02,Uchida.etal:04}.  This is more
typical of $K-L$ for the {\em nonaccreting} stars in Taurus that are
in the range $-0.05$-0.25 for similar spectral types (K7-M2)
\citep{Meyer.etal:97}.  Indeed, the inner disk of TW~Hya appears to be
almost completely cleared \citep[e.g.\
][]{Calvet.etal:02,Rettig.etal:04}.  Consequently there is very little
material in the form of small dust particles available to accrete and
replenish the gain-depleted gas.  Depletion is also in evidence in the
unexpectedly weak or absent Si lines in UV spectra
\citep{Valenti.etal:00,Herczeg.etal:02}, and in low Al and Si
abundances in the gas jet of TW~Hya inferred by
\citet{Lamzin.etal:04}.  This depletion is in qualitative accord with
the infrared detection of excess emission at 8-13$\mu $m attributable
to silicates, and the spectral energy distribution that indicates that
grain formation and coagulation into larger particles is well advanced
\citep{Weinberger.etal:02, Sitko.etal:00,Calvet.etal:02}.  Larger
grains would be expected to settle in the midplane of the
circumstellar disk \citep{Chiang.etal:01, D'Alessio.etal:99}.
\citet{Calvet.etal:02} speculate that the truncation of the outer disk
at $\sim 4$~AU inferred from the negligible near-infrared excess could
be a developing gap caused by a growing protoplanet.  In this
scenario, the depletion of metals and oxygen in the X-ray spectrum of
the shocked accreting gas would be the result of retention of these
elements in bodies of sufficient size that they are not significantly
affected by the drag of inwardly migrating gas that feeds the
accretion.



\section{Conclusions}


We have investigated the Ne/O abundance ratios of the accreting gas of
the T~Tauri stars TW~Hya and BP~Tau using high resolution X-ray
spectra of H-like and He-like resonance lines, and have compared these
to the same ratio for the coronal plasma of sample of
magnetically-active main-sequence and evolved field stars.  In the
case of BP~Tau, we find an Ne/O abundance ratio in good agreement with
that of the coronal plasma of field stars.  In contrast, the Ne/O
ratio for TW~Hya is higher than for the rest of the sample by a factor
of two.

We interpret these results in terms of a relative depletion of O in
the material accreting from the inner disk of TW~Hya as compared with
BP~Tau.  This can be attributed to the different, partially
age-related, evolutionary states of the disks of these stars.  In the
BP~Tau disk, dust is still present near the disk corotation radius and
can be ionized and accreted, re-releasing elements depleted onto
grains.  In the more evolved TW~Hya disk, evidence points to ongoing
coagulation of grains into much larger bodies, and possibly planets,
that can resist the drag of inward-migrating gas, and accreting gas is
consequently depleted of grain-forming elements.

These results demonstrate the utility of high resolution X-ray spectra
of CTTS for probing directly the chemical composition of the accreting
gas and the state of evolution of the very inner protoplanetary disk.

\acknowledgments

We thank the NASA AISRP for providing financial assistance for the
development of the PINTofALE package.  JJD was supported by NASA
contract NAS8-39073 to the {\em Chandra X-ray Center} during the
course of this research.  PT was supported by Chandra award number G03-4005A issued by CXC,
and SAO contract SV3-73016 to MIT for support of CXC, which is operated
by SAO for and on behalf of NASA under contracts NAS8-39073 and
NAS8-03060.



\newpage

\begin{deluxetable}{lcccccccc}
\tabletypesize{\footnotesize}
\tablecaption{Spectral line fluxes and derived Ne/O abundance ratios
(by number). \label{t:flx}}
\tablehead{
\colhead{} & \colhead{Spec.} & \colhead{d} &  
\colhead{} & \colhead{N$_{\rm H}$} &
   \multicolumn{3}{c}{Fluxes
[$10^{-13}$~erg~cm$^{-2}$~s$^{-1}$]\tablenotemark{e}} & \colhead{} \\ 
\colhead{Source} & \colhead{Type} & 
\colhead{[pc]} & \colhead{L$_{\rm X}/L_{\rm bol}$}  &
\colhead{[10$^{20}$~cm$^{-2}$]} &
\colhead{Ne~IX} &  \colhead{Ne~X} &
 \colhead{O~VIII} & \colhead{$A_{Ne}/A_{O}$}}
\startdata
TW~Hya      &  K7  &  56  &  $-3.05$  &  2.0\tablenotemark{b}   &  
2.29  $\pm$ 0.19\tablenotemark{a}   &   
1.52 $\pm$ 0.14\tablenotemark{a}  &  2.9  $\pm$
0.4\tablenotemark{a} &   0.87 $\pm$ 0.13 \\
       & \nodata & \nodata &  \nodata &  2.0\tablenotemark{b}  &  
3.64 $\pm$ 0.32\tablenotemark{b} & 1.09
$\pm$ 0.25\tablenotemark{b}  &  3.6   $\pm$ 0.3\tablenotemark{b}
   &   1.06 $\pm$ 0.13 \\
BP~Tau    &  K5  & 140  &  $-3.59$  & 10-20\tablenotemark{c} & 
0.108 $\pm$ 0.029\tablenotemark{c}  &   0.29
$\pm$ 0.04\tablenotemark{c}  &  0.28  $\pm$
0.03\tablenotemark{c} 
&   0.54 $\pm$ 0.12  \\
TWA 5 & M1.5 & 50 & -3.1  &  3.0\tablenotemark{d}  & 0.59 $\pm$ 0.13
\tablenotemark{d}  &  1.28 $\pm$ 
0.15\tablenotemark{d}  &  1.51 $\pm$ 0.9\tablenotemark{d}  & 0.52 \phs
\phm{0.13}

\enddata
\tablenotetext{a}{Based on {\it Chandra} MEG spectra (this work).}
\tablenotetext{b}{From \citet{Stelzer.Schmitt:04} based on XMM-Newton RGS
spectra.}
\tablenotetext{c}{From \citet{Schmitt.etal:05} based on XMM-Newton RGS
spectra.}
\tablenotetext{d}{From \citet{Argiroffi.etal:05} based on XMM-Newton RGS
spectra.}
\tablenotetext{e}{Fluxes have been corrected for the listed
intervening H column densities from \citet{Stelzer.Schmitt:04} for 
TW~Hya, \citet{Schmitt.etal:05} for BP Tau, and
\citet{Argiroffi.etal:05} for TWA~5.}
\end{deluxetable}
\tablenotemark{d}


\end{document}